\documentclass[sn-mathphys-num]{sn-jnl}% Math and Physical Sciences Numbered Reference Style 

\usepackage{graphicx}%
\usepackage{multirow}%
\usepackage{amsmath,amssymb,amsfonts}%
\usepackage{amsthm}%
\usepackage{mathrsfs}%
\usepackage[title]{appendix}%
\usepackage{xcolor}%
\usepackage{textcomp}%
\usepackage{manyfoot}%
\usepackage{booktabs}%
\usepackage{algorithm}%
\usepackage{algorithmicx}%
\usepackage{algpseudocode}%
\usepackage{listings}%
\usepackage{epstopdf}

\theoremstyle{thmstyleone}%
\theoremstyle{thmstyletwo}%

\theoremstyle{thmstylethree}%

\begin{document}

\title[Article Title]{Relationship between the shear moduli and defect-induced structural relaxation of high-entropy metallic glasses}

\author[1]{\fnm{Andrey} \sur{Makarov}}\email{a.s.makarov.vrn@gmail.com}

\author[1]{\fnm{Evgenia} \sur{Gonchrova}}\email{goncharova.evg@mail.ru}
\equalcont{These authors contributed equally to this work.}

\author[2]{\fnm{Jichao} \sur{Qiao}}\email{qjczy@nwpu.edu.cn}
\equalcont{These authors contributed equally to this work.}

\author[1]{\fnm{Roman  \sur{Konchakov}}}\email{konchakov@mail.ru}
\equalcont{These authors contributed equally to this work.}

\author*[1]{\fnm{Vitaly} \sur{Khonik}}\email{v.a.khonik@yandex.ru}
\equalcont{These authors contributed equally to this work.}

\affil[1]{\orgdiv{Department of General Physics}, \orgname{Voronezh State Pedagogical University}, \orgaddress{\street{Lenin Str. 86}, \city{Voronezh},   \postcode{394043}, \country{Russia}}}

\affil[2]{\orgdiv{School of Mechanics, Civil Engineering and Architecture}, \orgname{Northwestern Polytechnical University}, \orgaddress{\city{Xi'an}, \postcode{710072},  \country{China}}}

%\affil[3]{\orgdiv{Department}, \orgname{Organization}, \orgaddress{\street{Street}, \city{City}, \postcode{610101}, \state{State}, \country{Country}}}


\abstract{We performed  high-frequency shear modulus and calorimetry measurements on seven   high-entropy metallic glasses (HEMGs) in the initial, relaxed and crystalline states. It is shown that the shear modulus of HEMGs is intrinsically related with the concentration of defects responsible for structural relaxation. In the absence of structural relaxation, temperature coefficient of  shear modulus of glass equals to that of the maternal crystal. All found regularities are governed by a single equation.}




\maketitle

Many-component alloys can be characterized by the mixing entropy, which is a measure of how much disorder is present in an alloy. This quantity is defined as $\Delta S_{mix}=-R\sum\limits_{i=1}^nc_ilnc_i$
where $R$ is the universal gas constant, $c_i$ is the molar fraction of the \textit{i}-th element in the alloy and $n$ is the number of constituent elements. The alloys with a mixing entropy of $S_{mix} \geq 1.5R$ are referred to as high-entropy alloys \cite{YehAdvEngMater2004}. They were discovered in the early 2000s and consist of five or more metallic elements, each with an atomic percentage between 5\% and 35\%.  About a decade ago, it was discovered that high-entropy alloys can be solidified into a non-crystalline state, forming high-entropy metallic glasses (HEMGs) \cite{TakeuchiIntermetallics2011}. It was found that HEMGs exhibit enhanced thermal stability \cite{LuanNatComm2022}, which is manifested in higher activation energies for crystallization \cite{ChenIntermetallics2023}. Since the high-mixing-entropy state results in the low excess entropy of solid glass compared to the maternal crystal \cite{AfoninAPL2024}, HEMGs demonstrate lower atomic mobility \cite{ChenIntermetallics2023}, sluggish diffusion \cite{DuanPRL2022} and crystallization kinetics \cite{YangMaterResLett2018} as well as slow dynamics of homogeneous flow \cite{MakarovIntermet2024}.   It was suggested that HEMGs combine the properties of both metallic glasses and high-entropy crystalline alloys  \cite{ChenJALCOM2021}. It is important, therefore, to derive a thorough understanding of the fundamental physical properties of HEMGs. In particular, this applies to the shear elasticity, which is controlled by the instantaneous shear modulus.  Meanwhile, the instantaneous shear modulus (in practice, high-frequency shear modulus called simply shear modulus $G$ hereafter) controls the heights of barriers for local atomic rearrangements in different types of glasses, including HEMGs, and, therefore, constitutes their major physical parameter \cite{DyreRevModPhys2006}. 

Metallic glasses are produced by melting the maternal crystal and then rapidly cooling the melt.  One can raise an important question: is there any relationship between the properties of glass and its crystalline counterpart? This in full applies to the shear elasticity controlled by the shear moduli of glass and its maternal crystal. However, this issue is seldom discussed in the literature \cite{MitrofanovPhysStatSol2019}. 

On the other hand, metallic glasses are prone to spontaneous thermoactivated changes of their properties called structural relaxation. This phenomenon is often interpreted as a result of the changes in the system of defects -- local regions with low point symmetry \cite{ChengProgMaterSci2011}.  These defects are considered from very different viewpoints (e.g. Refs. \cite{ChengProgMaterSci2011,LiuSciRep2017}) and structural relaxation is most often interpreted in terms of changes of their concentration. 

Meanwhile, the relation between the defect system, the shear moduli of glass  and its maternal crystal is intrinsically built into the  Interstitialcy theory (IT), which was found to provide a quantitative understanding of different relaxation phenomena in metallic glasses (a review of the IT and its interpretation of  experimental data is given in  Ref.\cite{KobelevUFN2023}). The IT argues that the defect system of glass, the shear moduli of glass $G$ and its maternal crystal $\mu$ are interrelated as

\begin{equation}
G=\mu\;exp(-\alpha\beta c), \label{G}
\end{equation}
where $c$ is the concentration of defects similar to dumbbell interstitials in crystals and responsible for reduced shear modulus of glass (diaelastic effect), $\beta$ is dimensionless shear susceptibility and dimensionless $\alpha\approx 1$ \cite{KobelevUFN2023}. Eq.(\ref{G}) after differentiation can be rewritten as

\begin{equation}
 \beta \frac{dc(T)}{dT}=\frac{d}{dT} ln \frac{\mu(T)}{G(T)}. \label{dcdT}
 \end{equation} 
It  follows from this equation that if $c=const$ and, therefore, 
structural relaxation is absent then the  right-hand side must be zero.  In this case one can write down that 
\begin{equation}
\frac{d\;lnG(T)}{dT}=\frac{d\;ln\mu(T)}{dT}. \label{dG=dMu}
\end{equation}
This equation shows that temperature coefficients of the shear moduli in the glassy and crystalline states in the absence of structural relaxation should be equal, i.e. $G^{-1}dG/dT=\mu^{-1}d\mu/dT$.

On the other hand, in the presence of structural relaxation, the defect concentration $c$ below the glass transition temperature $T_g$ decreases, i.e. $dc/dT<0$, and, therefore, the derivative $\frac{d}{dT} ln \frac{\mu(T)}{G(T)}$ in Eq.(\ref{dcdT}) should be negative. Conversely, since $c$ increases   above  $T_g$, one should obtain the inequality $\frac{d}{dT} ln \frac{\mu(T)}{G(T)}>0$ in this case. All these predictions on the defect concentration, shear moduli of glass and maternal crystal as well as the validity of Eq.(\ref{G}) are  verified in  the present work.

We studied 7 HEMGs listed in Table 1 with the mixing entropies $1.42 \leq S_{mix}/R\leq 1.77$. The glasses were obtained by melt quenching into a copper mold and  X-ray verified to be fully amorphous. Differential scanning calorimetry (DSC) was performed with a Hitachi DSC 7020 instrument in high-purity nitrogen atmosphere using 50-70 mg samples. A crystallized sample of nearly the same mass was placed into the reference cell, so that the instrument measured the difference in the heat flow between the glassy and crystalline samples referred to as the differential heat flow $\Delta W$ hereafter.

Shear modulus was measured by the electromagnetic transformation method. In this method, transverse resonant vibrations ($f=400-600$ kHz) of a sample ($5\times 5\times 2$ mm$^3$) are produced due to Lorentz interaction of external magnetic field with surface current excited by a coil  \cite{VasilievUFN1983}.  The shear modulus was calculated as $G(T)=G_{rt}\frac{f^2(T)}{f^2_{rt}}$, where $f_{rt}$ and $G_{rt}$ are the vibration frequency and shear modulus at room temperature, respectively. The error of the determination of temperature changes of $G$ was about 5 ppm near room temperature and about 100 ppm near the glass transition temperature $T_g$. Room temperature shear modulus was determined using resonant ultrasound spectroscopy with a precision of 1-2\%.

\begin{table}[t]
%\begin{center}
%\footnotesize
\caption{\label{tab:table1} High-entropy metallic glasses under investigation, their mixing entropies and the derivatives $dlnG/dT$ and $dln\mu /dT$ calculated for temperature ranges with no structural relaxation.   } 
%\scriptsize
\begin{tabular}{p{3mm}|l|l|l|l}
\hline
\hline
No & Composition (at.\%)&$S_{mix}/R$&$dlnG/dT\; (\times 10^{-4}$ K$^{-1}$)&$dln\mu/dT\; (\times 10^{-4}$ K$^{-1}$)\\ 
\hline
\hline
1 & Zr$_{31.6}$Cu$_{37.8}$Hf$_{13.4}$Al$_{8.7}$Ag$_{8.4}$&1.42&$-2.69\pm 0.03$& $-2.54\pm 0.04$\\
2 & Zr$_{31}$Ti$_{27}$Be$_{26}$Cu$_{10}$Ni$_{6}$ & 1.46 & $-2.44\pm 0.05$ & $-2.26\pm 0.04$\\
3 & (Ti$_{37.31}$Zr$_{22.75}$Be$_{26.39}$Al$_{4.59}$Cu$_9)_{94}$Co$_{6}$& 1.56 &$-2.34\pm 0.05$ & $-2.11\pm0.05$ \\
4 & Ti$_{20}$Zr$_{20}$Hf$_{20}$Be$_{20}$Cu$_{20}$  & 1.61 & $-2.57\pm0.08$ & $-2.32\pm 0.03$\\
5 & Ti$_{20}$Zr$_{20}$Hf$_{20}$Be$_{20}$Ni$_{20}$&1.61& $-2.62\pm 0.07$ & $-2.47\pm 0.08$ \\
6 & Zr$_{35}$Cu$_{25}$Hf$_{13}$Al$_{11}$Ag$_{8}$Ni$_{8}$ &1.63& $-2.46\pm 0.03$ & $-2.36\pm 0.03$\\
7 & Zr$_{35}$Hf$_{17.5}$Al$_{12.5}$Ni$_{12}$Cu$_{10}$Co$_{7.5}$Ti$_{5.5}$ & 1.77 & $-2.29\pm 0.06$ & $-2.28\pm 0.04$  \\
\hline
\hline
\end{tabular}
%\end{ruledtabular}
%\end{center}
\end{table}
 
\begin{figure}[t]
\centering
\includegraphics[scale=0.85]{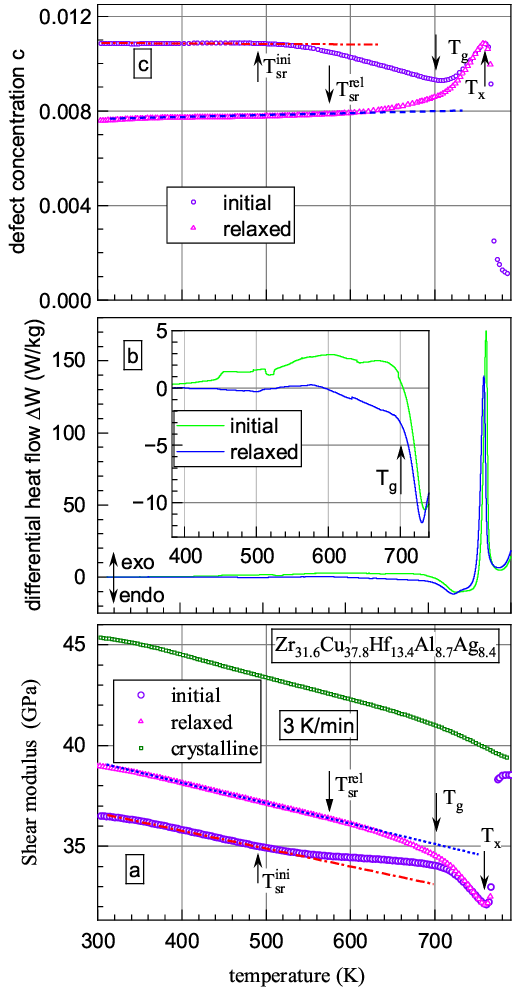}
\caption[*]{\label{Fig1.eps} Temperature dependences of the shear modulus (a), differential heat flow $\Delta W$ (b) and defect concentration (c) $c$ of glassy Zr$_{31.6}$Cu$_{37.8}$Hf$_{13.4}$Al$_{8.7}$Ag$_{8.4}$ in the initial and relaxed states. The defect concentration $c$ is calculated using Eq.(\ref{G}) with $\beta=20$. Temperatures of structural relaxation onset in the initial state ($T_{sr}^{ini}$) and relaxed state ($T_{sr}^{rel}$) together with the calorimetric glass transition temperature $T_g$ and crystallization onset temperature $T_x$ are indicated. The sold lines give corresponding linear approximations.}   
\label{Fig1}
\end{figure}

An example of shear modulus measurements is presented in Fig.\ref{Fig1}(a), which gives temperature dependences of $G$ in the initial state, after relaxation performed by heating into the supercooled liquid state (i.e. into the range $T_g<T<T_x$, where $T_x$ is the crystallization onset temperatures) and in fully crystalline (maternal) state. In  the initial state, $G$ first linearly decreases with temperature due to the anharmonicity, next starts to increase  due to exothermal  relaxation (see  DSC scan in Fig.\ref{Fig1}(b)) at a temperature $T_{sr}^{ini}\approx 490$ K, then begins to fall above $T_g\approx 701$ K and finally rapidly increases upon crystallization onset at $T_x\approx 759$ K. In the relaxed state, room-temperature value of $G$ is increased by $\approx 7\%$ and  the upward $G$-rise because of exothermal structural relaxation  is absent. Instead, heating results in endothermal relaxation (see Fig.\ref{Fig1}(b)) and shear modulus starts to fall below purely anharmonic decrease at a temperature $T_{sr}^{rel}\approx 575$ K. Above $T_g$, temperature behaviour of relaxed sample repeats that of the initial specimen. Crystallization leads to an  increase   of $G$ by $\approx 24\%$ and $G$ smoothly decreases with temperature upon heating. 

\begin{figure}[t]
\centering
\includegraphics[scale=1.1]{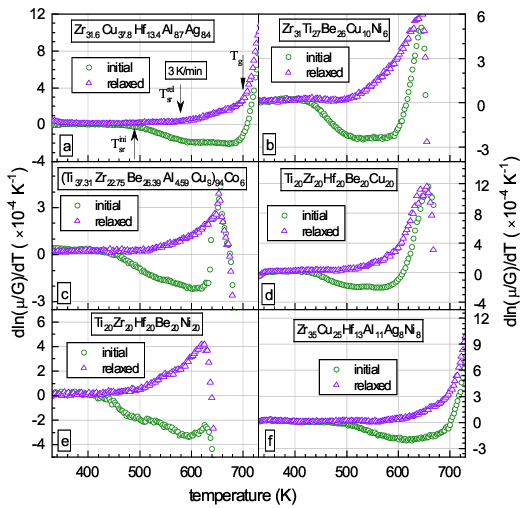}
\caption[*]{\label{Fig2.eps} Temperature dependences of the derivative $dln(\mu/G)/dT$ for indicated HEMGs in the initial and relaxed states.}  
\label{Fig2}
\end{figure} 

Figure \ref{Fig1}(c) gives the defect concentration $c$ of the same glass in the initial and relaxed states as a function of temperature calculated using Eq.(\ref{G}) with $G(T)$ and $\mu (T)$ dependences shown in Fig.\ref{Fig1}(a) and a typical shear susceptibility $\beta=20$. In the initial state, $c\approx 0.012$ and is nearly constant up to the temperature of exothermal structural relaxation onset $T_{sr}^{ini}$, next $c$ decreases by about one tenth near $T_g$ and rapidly increases above $T_g$. After relaxation, $c$ is decreased by about one third at room temperature and remains constant upon heating up to the beginning of endothermal structural relaxation at $T_{sr}^{rel}$, which results in an increase of the shear modulus up to $T\approx T_g$. Above $T_g$, temperature dependences of $G$ in the initial and relaxed states are close to each other. Similar data were obtained for other HEMGs. 

\begin{figure}[t]
\centering
\includegraphics[scale=0.85]{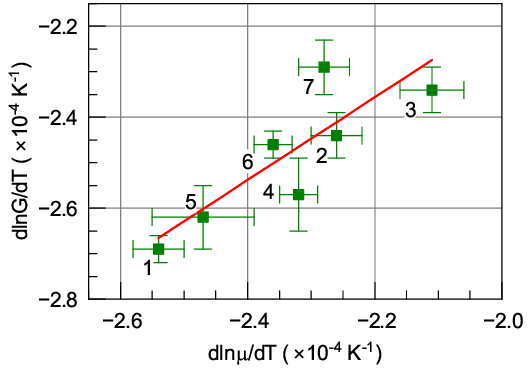}
\caption[*]{\label{Fig2.eps} Interdependence of the derivatives $dlnG/dT$ and $dln\mu /dT$ for the glassy and crystalline states. The former derivative is calculated for low temperature regions with no structural relaxation (i.e. below the temperatures denoted as $T_{sr}^{ini}$ in Fig.\ref{Fig1}(a)). The numbers give HEMGs compositions according to Table 1.  The slope of the linear fit is $0.91\pm 0.18$.   }  
\label{Fig3}
\end{figure} 

We can now check the validity of Eq.(\ref{dcdT}). This is done in Fig.\ref{Fig2}, which gives  the derivatives $\frac{d}{dT} ln \frac{\mu(T)}{G(T)}$  for six HEMGs  in the initial and relaxed states. The plots are quite similar and can be commented using the data on Zr$_{31.6}$Cu$_{37.8}$Hf$_{13.4}$Al$_{8.7}$Ag$_{8.4}$ (Fig.\ref{Fig2}(a)). Below the onset of exothermal relaxation of the initial glass at $T_{sr}^{ini}\approx 490$ K, this derivative is zero. Above this temperature, the derivative becomes negative reflecting a decrease of the defect concentration (see Fig.\ref{Fig1}(c)), in accordance with Eq.(\ref{dcdT}). Above  $T_g$, the defect concentration rapidly increases due to endothermal relaxation in the supercooled liquid state (Fig.\ref{Fig1}(b)) so that $\frac{d}{dT} ln \frac{\mu(T)}{G(T)}$ sharply increases with temperature, in line with Eq.(\ref{dcdT}). After relaxation, $\frac{d}{dT} ln \frac{\mu(T)}{G(T)}$ remains close to zero up the beginning of endothermal relaxation at $T_{sr}^{rel}\approx 575$ K (Fig.\ref{Fig2}(b)) and increases at higher temperatures that follows the growing defect concentration $c$. Entering the supercooled liquid state at $T>T_g$ results in a strong endothermal flow (Fig.\ref{Fig2}(b)) leading to a rapid increase of both  $c$ and $\frac{d}{dT} ln \frac{\mu(T)}{G(T)}$, which is reflected by Eq.(\ref{dcdT}). Similar interconnection between the heat effects, defect concentration $c$ and derivative $\frac{d}{dT} ln \frac{\mu(T)}{G(T)}$  is observed for other HEMGs.  

In the absence of structural relaxation (i.e. if $dc/dT=0$), Eq.(\ref{dG=dMu}) should be obeyed.  This prediction is verified in Fig.\ref{Fig3}, which gives the interdepedence between the derivatives $dlnG/dT$ and $dln\mu/dT$, which are calculated for the temperature ranges of 300 K$<T<T_{sr}^{ini}$ (in these ranges $G(T)$ and $\mu (T)$ are linear in temperature) for all indicated HEMGs. It is seen that the slope of the linear fit for this interdependence is close to unity that means the equality of  temperature coefficients $\frac{dG}{GdT}$ and $\frac{d\mu}{\mu dT}$ of the shear moduli for glass and maternal crystal in the absence of structural relaxation. At that, any dependence of these derivatives on the mixing entropy $\Delta S_{mix}$ is not detected. 

One can conclude, thus, that the shear moduli $G$ of HEMGs are intrinsically related with the defect concentration $c$: relaxation increase of  $G$ originates from an exothermal decrease of $c$. Conversely, a decrease of $G$ due to structural relaxation takes place because of endothermal rise of the defect concentration. The latter process is strongly pronounced in the supercooled liquid state above $T_g$. If thermoactivated structural relaxation is absent, then temperature coefficients of the shear moduli in the glassy and crystalline states are equal.         

All aforementioned regularities originate from  Eq. (\ref{G}), which constitutes the main hypothesis of the Interstitialcy theory \cite{KobelevUFN2023}. The present experiments and data analysis, therefore, provide further support for this theory.

\bmhead{Acknowledgements}

This work was supported by the Russian Science Foundation under the Grant No. 23-12-00162.

\bmhead{Conflict of interests}
The authors declare no competing interests.

\end{document}